\documentclass[final,3p,times]{elsarticle}
\usepackage{graphicx}
\usepackage{dcolumn}
\usepackage{bbm}
\usepackage{textcomp}
\usepackage{color}
\usepackage{gensymb}
\usepackage{wasysym}
\usepackage{amssymb}

\begin{document}

\def\nuc#1#2{${}^{#1}$#2}
\def\BBz{0$\nu\beta\beta$}
\def\BBt{2$\nu\beta\beta$}
\def\BB{$\beta\beta$}
\def\Tz{$T^{0\nu}_{1/2}$}
\def\Tt{$T^{2\nu}_{1/2}$}
\def\QBB{Q$_{\beta\beta}$}
\def\mBB{$\left < \mbox{m}_{\beta\beta} \right >$}

\title{Solubility, Light Output and Energy Resolution Studies of Molybdenum-Loaded Liquid Scintillators}
\author[label1,label2,label3]{V.M. Gehman}\ead{vmg@lanl.gov}
\author[label2,label3]{P.J. Doe}
\author[label2,label3]{R.G.H. Robertson}
\author[label2,label3]{D.I. Will}
\author[label4,label5]{H. Ejiri}
\author[label6]{R. Hazama}
\address[label1]{Los Alamos National Laboratory, Los Alamos, NM 87545}
\address[label2]{Center for Experimental Nuclear Physics and Astrophysics, University of Washington, Seattle, WA 98195}
\address[label3]{Department of Physics, University of Washington, Seattle, WA 98195}
\address[label4]{Research Center for Nuclear Physics, Osaka University, Osaka 567-0047, Japan}
\address[label5]{Nuclear Science, Czech Technical University, Brehov\'{a}, Prague, Czech Republic}
\address[label6]{Hiroshima University, Higashi Hiroshima, Hiroshima 739-8527}

\begin{abstract}
The search for neutrinoless double-beta decay is an important part of the global neutrino physics program.  One double-beta decay isotope currently under investigation is \nuc{100}{Mo}.  In this article, we discuss the results of a feasibility study investigating the use of molybdenum-loaded liquid scintillator.  A large, molybdenum-loaded liquid scintillator detector is one potential design for a low-background, internal-source \BBz\ search with \nuc{100}{Mo}.  The program outlined in this article included the selection of a solute containing molybdenum, a scintillating solvent and the evaluation of the mixture's performance as a radiation detector.
\end{abstract}

\maketitle

\section{Double Beta Decay of \nuc{100}{Mo}}\label{sec:Intro}
The study of neutrino physics, particularly of neutrinoless double-beta decay (\BBz) presents an exciting opportunity in the search for physics beyond the standard model and is well-motivated in the literature \cite{Avignone08, Ejiri05, Barabash04}.  One isotope of experimental interest in the search for \BBz\ is \nuc{100}{Mo}.  \nuc{100}{Mo} is currently under investigation by the MOON (Molybdenum Observatory Of Neutrinos)\cite{Ejiri08, Ejiri00, Nakamura07, Ejiri2004} and NEMO (Neutrino Ettore Majorana Observatory)\cite{NEMOBG, NEMO150Nd, NEMO100Mo, NEMO3Detector} projects.  All \BBz\ searches examine the Majorana nature of the neutrino (that is, whether the neutrino is its own anti-particle) as well as the absolute neutrino mass scale and spectrum.  The experimental signature of most \BBz\ searches is a pair of electrons in the final state of the decay whose total kinetic energy equals the endpoint of the two-neutrino double-beta decay (\BBt) spectrum, \QBB\ (3034.4 keV for \nuc{100}{Mo}\cite{QValue}).  \nuc{100}{Mo} is a particularly attractive \BBz\ candidate for several reasons.  First, its \QBB\ is both above most $\gamma$ ray related backgrounds and below most $\alpha$ decay energies.  Second, \nuc{100}{Mo} has a rather large natural abundance at 9.6\%.  Both MOON and NEMO look for this signature by interleaving source foils/films of \BB\ isotope between radiation detectors.  MOON is focused on \nuc{100}{Mo} (with a secondary effort on \nuc{82}{Se}), and NEMO looks at several isotopes in addition to \nuc{100}{Mo}.  This article will focus on attempts to load molybdenum into organic liquid scintillator as an alternative to the foil/tracking detector design.

We begin by discussing the motivation for the interest in liquid scintillators for \BB\ search experiments in section \ref{sec:WhyLS}.  We  then move on to molybdenum solubility and stability studies for different organometallic solutes in different scintillating solvents in Section \ref{sec:SolStabStudy}.  We next discuss the evaluation of the scintillation performance (light output and energy resolution) of a subset of the best performing solute/solvent combinations in Section \ref{sec:ScintLight}.  Last, we draw some conclusions and outline potential extensions to this program in Section \ref{sec:Concl}.

\section{Molybdenum Loaded Liquid Scintillators}\label{sec:WhyLS}
As discussed in Section \ref{sec:Intro}, most of the current experimental interest in \BB\ of \nuc{100}{Mo} involves sandwiching thin layers of Molybdenum foil/film between radiation detectors, such as large scintillator plates or some kind of tracking detector.  This design has the advantage of being able to measure tracks from the \BB\ electrons individually, and provides extra kinematic information about each event ({\it e.g.} individual electron energies and angular distribution), allowing a potential for greater physics reach and background rejection.  Such detectors are however quite complex and difficult to build and characterize.  

A conceptually simpler detector design (and therefore one easier to scale up to very large exposures of \BB\ isotope) is a ``Fiorini-Style'' internal source experiment, in which the \BB\ source and the radiation detector are the same physical objects\cite{Fiorini}.  One way to build such an internal source \nuc{100}{Mo} \BB\ experiment would be to dissolve a compound containing molybdenum into a scintillating solvent.  Organic liquid scintillators have a long history in physics experiments.  They are very well characterized and have excellent light yield.  The liquid phase of the scintillator also allows for careful refining and therefore, very low residual radioactivity levels and excellent optical properties.  Furthermore, housing the scintillator and PMT array in a large spherical tank would minimize the ratio of surface area to volume, leaving less area for potential contamination.  A spherically symmetric detector geometry would also simplify light collection and event reconstruction.  The SNO+ collaboration is pursuing a detector design of this kind in their search for \BBz\ in \nuc{150}{Nd}\cite{SNOPlusICHEP2008, SNOPlusMEDEX2007, SNOPlusErice2005}.  This design too, has its own problems to overcome.  Most notable of these is finding a combination scintillator/solvent and molybdenum-containing solute capable of maintaining high concentration (to facilitate large \nuc{100}{Mo} exposure) and optical attenuation lengths of many meters (the characteristic path length of optical photons through a potential detector).  Scintillation light output and energy resolution are also quite important because \nuc{100}{Mo} has a rather fast \BBt\ rate compared to other \BB\ isotopes.  The high-energy portion of the \BBt\ continuum can leak into the region of interest for the \BBz\ peak and represents an irreducible background for any \BBz\ search.  This \BBt\ background is only mitigated by fielding a detector with sufficiently fine energy resolution to either separate the \BBz\ peak from the \BBt\ continuum or fit for a spectral distortion of the \BBt\ spectrum at \QBB.  Efforts to address light output and energy resolution are discussed below.

\section{Solubility and Stability Studies}\label{sec:SolStabStudy}
The program to study the solubility and stability initially proceeded along two paths.  The first explored the simple dissolution of organometallic compounds containing molybdenum in commercial liquid scintillator.  The initial work here used hexacarbonylmolybdenum, Mo(CO)$_{6}$ in BC-505 (a commercially available scintillator from Bicron based on pseudocumene, PC).  Mo(CO)$_{6}$ was of particular interest because the carbonyl process excludes uranium and other potential radioactive contamination.  The Sudbury Neutrino Observatory (SNO) collaboration used this technique to obtain high-purity nickel for the proportional counters deployed in the third phase of the SNO experiment\cite{SNONCD}.  This method yielded modest molybdenum solubilities (of order 0.1--0.5\% by mass) and good optical clarity.  The second approach paralleled some of the work done in support of the LENS collaboration's development of Yb and In-loaded liquid scintillators\cite{LENS2001, RELoadedScint}.  This technique incorporated Ammonium Isovalerate (C$_{5}$H$_{13}$NO$_{2}$) to increase the solubility of molybdenum chloride in BC-505.  None of the solutions prepared in this way were optically clear (most had a color and consistency similar to that of black ink).  The lack of optical clarity made these options unfeasible for a large scintillation detector.  Therefore, no quantitative solubility measurements were made.

Because the efforts on direct dissolution were more promising, we then expanded on it in the next phase of the program.  We focused on five liquid scintillators as solvents: benzene (C$_{6}$H$_{6}$, BZ), toluene (C$_{7}$H$_{8}$, TL), xylene (C$_{8}$H$_{10}$, XL, actually a mixture of the three different xylene isomers), pseu\-doc\-umene (C$_{9}$H$_{12}$), and 1-meth\-yl\-naphtha\-lene (C$_{11}$H$_{10}$, MN).  For molybdenum-containing solutes, we attempted the dissolution of: molybdenum car\-bon\-yl, bis\-(a\-ce\-tyl\-ace\-ton\-ato)-di\-oxo\-mol\-yb\-den\-um ([CH$_{3}$COCH=C(O-)CH$_{3}$]2MoO$_{2}$, BAADOM), mol\-yb\-den\-um tri\-oxide (MoO$_{3}$, which while not an organic compound, turned out to be slightly soluble, inexpensive, and comparatively non-toxic), and several ammonium molybdates (compounds including molybdenum-based poly-atomic ions): amm\-on\-ium mol\-yb\-date tetra\-hy\-drate (H$_{24}$Mo$_{7}$N$_{6}$O$_{24}4$H$_{2}$O, AMTH), amm\-on\-ium di\-mol\-yb\-date (NH$_{4}$2Mo$_{2}$O$_{7}$, ADM) and amm\-on\-ium mol\-yb\-date (NH$_{4}$Mo$_{2}$O$_{7}$, AM).  We prepared supersaturated solutions of each solute in each solvent, and then spun them in a centrifuge to remove any non-dissolved precipitate from the mixture, leaving dissolutions at saturation.  We measured the concentration of Mo for each solute/solvent pair using flame atomic absorption (FAA) techniques.  The results of the solubility tests are presented in Table \ref{tab:MoConc}.
\begin{table}[htdp]
\caption{Saturation concentration (\% Mo by mass) as measured by FAA for different solute/solvent combinations.  See the text for more detail as well as definitions for each acronym.}
\begin{center}
\begin{tabular}{c|c|c|c|c|c}
\hline\hline
             &\multicolumn{4}{c}{Solvent} \\
Solute       &BZ    &TL    &XL    &PC    &MN \\
\hline
Mo(CO)$_{6}$ &0.54  &0.53  &0.44  &0.34  &0.21 \\
BAADOM       &0.65  &0.34  &0.23  &0.09  &0.73 \\
MoO$_{3}$    &0.007 &0.005 &0.002 &0.003 &0.026 \\
AMTH         &0.006 &0.004 &0.002 &0.002 &0.004 \\
ADM          &0.002 &0.002 &0.002 &0.002 &0.006 \\
AM           &0.002 &0.002 &0.002 &0     &0.004 \\
\hline\hline
\end{tabular}
\end{center}
\label{tab:MoConc}
\end{table}

During subsequent parts of this program, we observed the solutions over time to check for stability.  Of particular concern was optical degradation and slow precipitation of molybdenum out of the solution.  Nearly all samples with a high concentration of Mo (including the standards used for calibrating the FAA tests) acquired a slight blue tint over time.  We surmise that this is the product of Mo in the solution oxidizing or hydrolyzing due to contamination (probably air leaks).  Further work would be required to confirm this, as well as to quantitatively measure their attenuation lengths over time.  The highest molybdenum solubility levels were obtained for Mo(CO)$_{6}$ and BAADOM in all five solvents.  Optical degradation was much stronger in the BAADOM solutions.  BAADOM is also far more toxic than Mo(CO)$_{6}$.  These two facts, combined with our prior experience with the radiological cleanliness of carbonyl compounds led us to focus on the Mo(CO)$_{6}$ for the light output studies.  We achieved similar saturation concentrations for Mo(CO)$_{6}$ in BZ and TL.  Again to avoid toxicity issues associated with handling benzene, we opted to study the light output of Mo(CO)$_{6}$ in toluene in more detail.

\section{Scintillation Performance}\label{sec:ScintLight}
To better quantify the light output of the metal-loaded organic liquid scintillator, we required a standard scintillation candle.  For this, we chose two industry-standard liquid scintillators: BC-505, and a toluene-based liquid scintillator from Aldrich (catalog number: 327123-250ML).  We were also concerned that the fluors added to liquid scintillators would change the solubility of the Mo compounds in our tests.  We wanted the ability to add fluors to the scintillator mixtures after Mo-loading.  This allowed us to check for optical degradation and solid precipitates in the presence and absence of the fluors.  We mixed non-molybdenum-loaded liquid scintillators based on each of the five scintillators discussed above.  We also tested the light output of these ``home-brewed'' liquid scintillators, comparing them to the commercial ones.  As fluors, 2,5-Diphenyloxazole (PPO) and 1,4-Bis(5-phenyloxazol-2-yl)benzene (POPOP) were added to the base solvents in accordance with the guidelines in Reference \cite[Chapters 3 and 10]{Birks}.  All told, we tested the light output of eight different liquid scintillators: our two standard candles, five home-brewed scintillators (one from each of the five organic scintillators), and one Mo-loaded toluene solution.

\begin{figure}
\begin{center}
\includegraphics[angle=0,width=10cm]{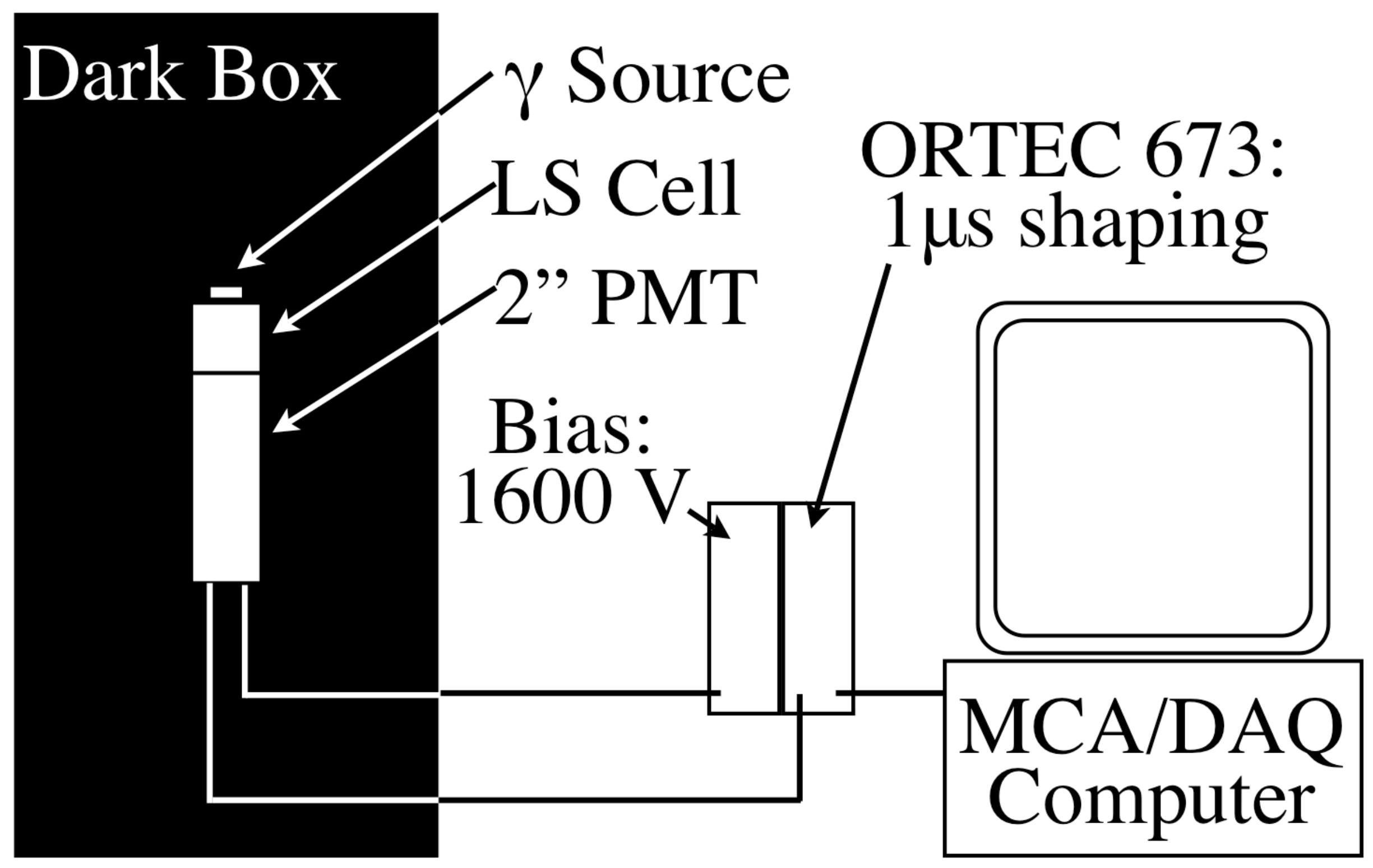}
\caption{\label{fig:PMTSetup} Scintillation counting setup schematic showing: the dark box, liquid scintillator cell, two-inch photomultiplier tube and supporting electronics.}
\end{center}
\end{figure}

In preparation for each run, we sparged each scintillator with high-purity N$_{2}$ gas for roughly thirty minutes to remove residual dissolved oxygen.  The scintillator was then transferred to a 5-cm diameter cylindrical test cell, filled to a depth of 2.5 cm.  This gave a total scintillator volume of 49 mL.  A 5-cm PMT was then optically coupled to the bottom of the cell.  The tube/test cell optical module was wrapped in aluminized mylar, sealed with black tape and placed in a light-tight box.  Each of two $\gamma$-ray sources was then, in turn, placed on the top-center of the test cell.  The output of the PMT was fed through an ORTEC 673 spectroscopy amplifier into a multi-channel analyzer (MCA) card in the data acquisition computer.  The counting setup is shown schematically in Figure \ref{fig:PMTSetup}.
The spectrum files captured by the MCA were analyzed in the ROOT analysis framework.  

\begin{figure}
\begin{center}
\includegraphics[angle=90,width=12cm]{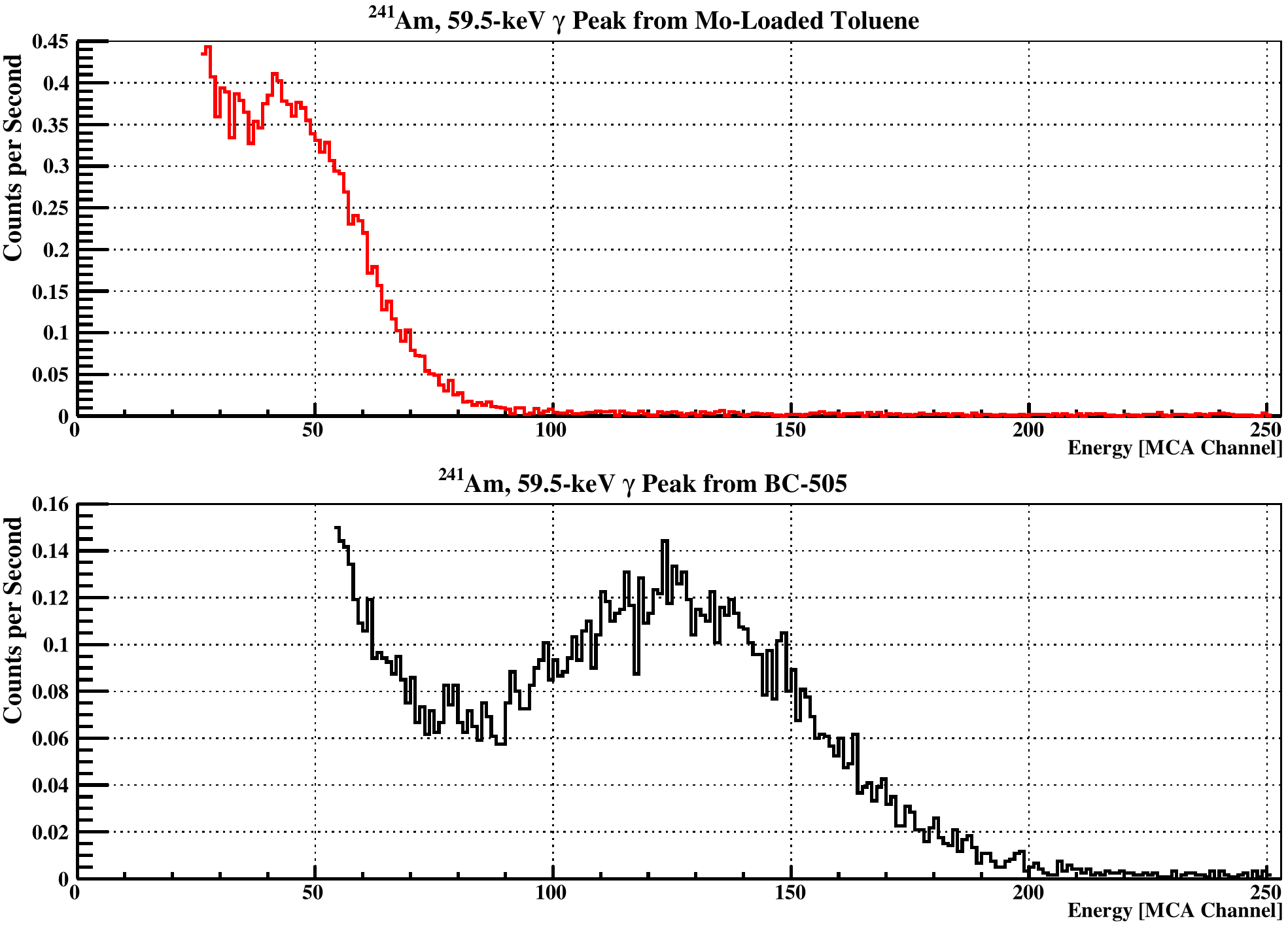}
\caption{\label{fig:AmSpec} 59.5-keV photo-peak from \nuc{241}{Am} acquired with Mo-loaded TL (top, peak at $\sim 40$ MCA channels) and BC-505 (bottom, peak at $\sim 120$ MCA channels).}
\end{center}
\end{figure}

As stated above, we acquired spectra with our eight liquid scintillator mixtures from two sources: \nuc{137}{Cs} and \nuc{241}{Am}.  We were able to capture the photo-peak from the 59.5-keV $\gamma$-ray from \nuc{241}{Am}.  We fit a simple Gaussian plus a flat background to the the 59.5-keV photo-peak in the \nuc{241}{Am} data.  Examples of the \nuc{241}{Am} photo-peak acquired with BC-505 and Mo-loaded TL are presented in Figure \ref{fig:AmSpec}.

\begin{figure}
\begin{center}
\includegraphics[angle=90,width=12cm]{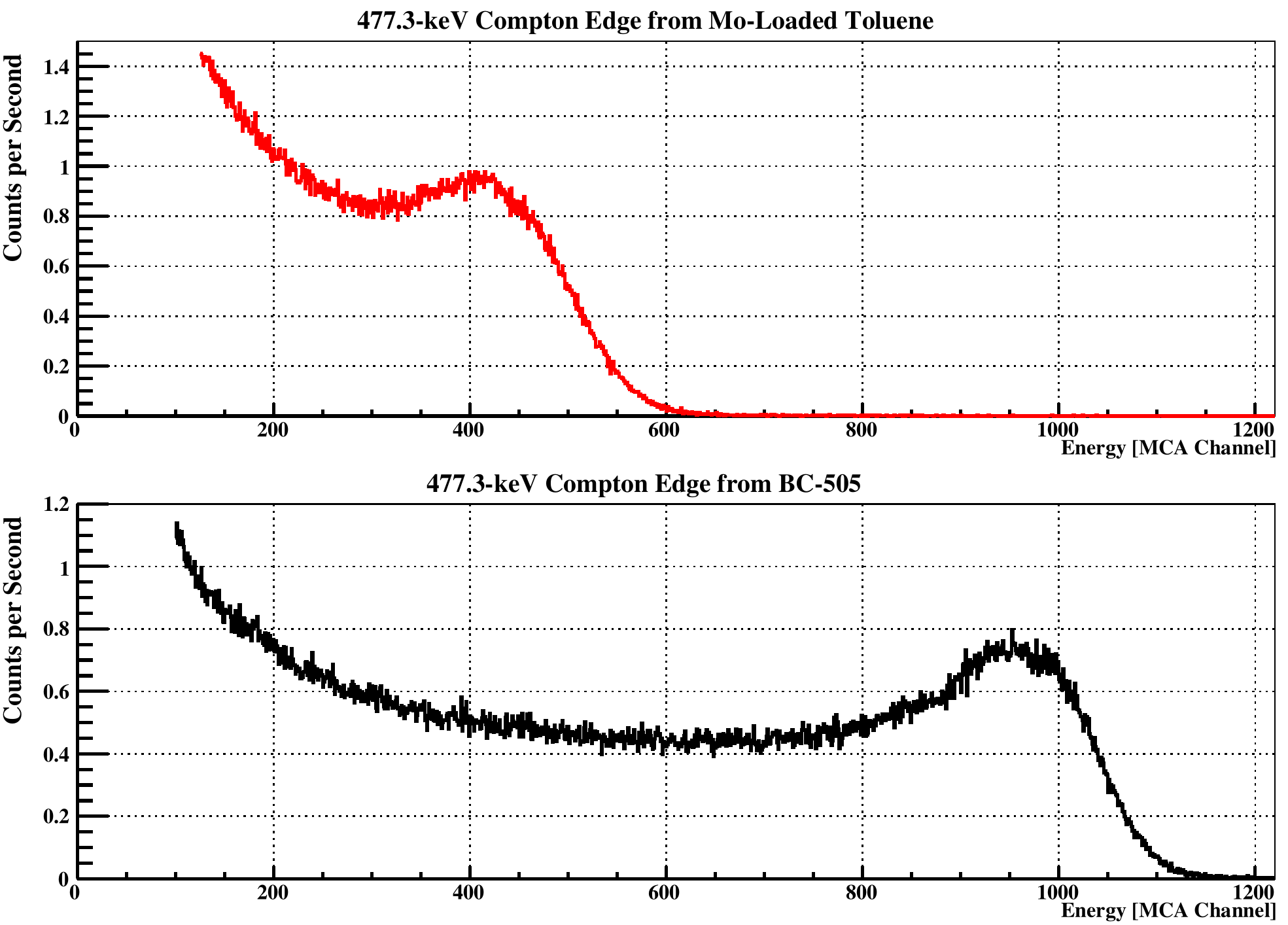}
\caption{\label{fig:CsSpec} 477.3-keV Compton edge from \nuc{137}{Cs} acquired with Mo-loaded TL (top, edge at $\sim 498$ MCA channels) and BC-505 (bottom, edge at $\sim 1038$ MCA channels).}
\end{center}
\end{figure}

We were only able to measure Compton scattering spectra from the \nuc{137}{Cs} source because its $\gamma$-ray line energy (661.7 keV) was too high to capture a full-energy photo-peak in our small scintillator volume.  The photo-electric process is much smaller than the Compton scattering process in our low-Z scintillator at these energies and the Compton-scattered photons escaped the scintillator without fully depositing their energy.   Analyzing the Compton spectra was more complicated than the photo-peaks.  An ``ideal'' Compton spectrum ({\it i.e.} one measured using a detector with perfect resolution) would have a sharp cutoff corresponding to the maximum energy of an electron scattered by an incoming $\gamma$ ray of a specific energy (Reference \cite{Nachtmann} provides an excellent discussion of this).  This cutoff occurs at 477.3 keV for the 661.7-keV $\gamma$-ray from \nuc{137}{Cs}.  Any real data will, of course have a less than perfect cutoff, corresponding to the finite energy resolution of the detector used to capture it as well as multiple Compton scatters.  Examples of the \nuc{137}{Cs} Compton edge spectra acquired with BC-505 and Mo-loaded TL are presented in Figure \ref{fig:CsSpec}.
To understand the Compton edge spectra, we employed a formalism outlined in Reference \cite{ELEGANTSV,Ejiri2001}.  We used a simple Monte Carlo to generate simulated Compton spectra with finite fractional energy resolutions ($\sigma$/E) ranging from two to ten percent.  We then located the energy of the Compton edge and and the Compton edge plus one sigma in each simulated spectrum and computed the intensity in that bin relative to the maximum near the Compton edge.  We repeated this process one thousand times for each fractional resolution, and stored these relative heights for each simulated spectrum.  The relative height histograms from our simulations allowed us to estimate the uncertainties (we simply used the RMS width of the histogram) as well the central value for both the Compton edge and the energy resolution in our real data.  We found that in the \nuc{137}{Cs} simulations, the Compton edge was where the spectrum dropped to $55.5 \pm 7.3$\% of its maximum value.  The Compton edge plus one sigma was where the spectrum fell to $17.4 \pm 3.8$\% maximum.  The one sigma energy resolution in our data is the difference of these two uncalibrated energies.  We can recover an energy resolution in keV by comparing the ratio of the uncalibrated energy resolution and Compton edge to the known energy of that Compton edge.

We tabulate the results of the scintillation light studies described above in Tables \ref{tab:AmScintOutput} and \ref{tab:CsScintOutput} for the two standard candles, our home made TL scintillator and the Mo-loaded TL scintillator.  The light yield for each scintillator quantified by the location of the Compton edge (or photo-peak), and is presented both as the MCA channel in the uncalibrated spectrum and as a fraction of the Compton edge (or photo-peak) obtained for BC-505.  We found the scintillation light yield for Mo-loaded TL to be roughly half what that of BC-505.  We quantify the energy resolution of each scintillator as $\sigma$, not FWHM, and present it in absolute MCA channels and as a fraction of the Compton edge (or photo-peak) location.
\begin{table}[ht]
\begin{center}
\caption{\label{tab:AmScintOutput} Scintillation output results for \nuc{241}{Am}, $E_{\gamma}$ = 59.5 keV.}
\renewcommand{\arraystretch}{1.0}
\begin{tabular}{c|c|c|c|c}
\hline\hline
Solution           &Peak Energy (MCA Ch) &Energy (\% BC-505) &$\sigma$ (MCA Ch) &$\sigma$ (\% Peak)\\
\hline
BC-505             &$122.5 \pm 29.7$     &$100.0 \pm 34.3$   &$30.4 \pm 15.9$   &$24.8 \pm 14.3$\\
Commercial TL      &$104.3 \pm 22.1$     &$85.2 \pm 27.4$    &$24.7 \pm 15.3$   &$23.7 \pm 15.5$\\
Home made TL       &$78.9 \pm 15.8$      &$64.4 \pm 20.3$    &$21.0 \pm 13.4$   &$26.6 \pm 17.8$\\
Mo(CO)$_{6}$ in TL &$42.6 \pm 16.0$      &$34.8 \pm 15.5$    &$16.1 \pm 9.7$    &$37.7 \pm 26.8$\\
\hline\hline
\end{tabular}
\end{center}
\end{table}
\begin{table}[ht]
\begin{center}
\caption{\label{tab:CsScintOutput} Scintillation output results for \nuc{137}{Cs}.  $E_{\gamma}$ = 661.7 keV.  Compton Edge = 477.3 keV}
\renewcommand{\arraystretch}{1.2}
\begin{tabular}{c|c|c|c|c}
\hline\hline
Solution           &Cutoff (MCA Ch)  &Cutoff (\% BC-505)    &$\sigma$ (MCA Ch) &$\sigma$ (\% cutoff)\\
\hline
BC-505             &$1038^{+7}_{-8}$ &$100.0^{+1.0}_{-1.1}$ &$45^{+8}_{-7}$    &$4.3^{+0.8}_{-0.7}$\\
Commercial TL      &$1022^{+7}_{-5}$ &$98.5 \pm 0.9$        &$52^{+8}_{-11}$   &$5.1^{+0.8}_{-1.1}$\\
Home made TL       &$778^{+6}_{-15}$ &$75.0^{+0.8}_{-1.6}$  &$68^{+4}_{-9}$    &$8.7^{+0.5}_{-1.2}$\\
Mo(CO)$_{6}$ in TL &$498^{+9}_{-10}$ &$48.0^{+0.9}_{-1.0}$  &$54^{+10}_{-8}$   &$10.8^{+2.0}_{-1.6}$\\
\hline\hline
\end{tabular}
\end{center}
\end{table}

We characterized the energy resolution of our Mo-loaded TL at 59.5 and 477.3 keV, but for \BBz\ in \nuc{100}{Mo}, the energy resolution at \QBB\ is much more important.  To estimate the energy resolution at \QBB, we examine the simulations from Reference \cite{ELEGANTSV,Ejiri2001}.  The energy resolution (FWHM) reported in Reference \cite{ELEGANTSV,Ejiri2001} was 57 keV at 477.3 keV and 167 keV at \QBB.  The ratio of these two resolutions is 2.9.  We can then apply this same scale factor to the energy resolution from our data at 477.3 keV to estimate the resolution of our Mo-loaded scintillator at \QBB:
\begin{equation}
\frac{\sigma(\mbox{Q}_{\beta\beta})}{\mbox{Q}_{\beta\beta}} = 2.9 \times \left ( 0.108^{+0.02}_{-0.016} \times \frac{477.3\ \mbox{keV}}{\mbox{Q}_{\beta\beta}} \right ) = 4.9^{+0.9}_{-0.7}\%.
\end{equation}
The factor of 0.108 comes from the 10.8\% fractional energy resolution of our Mo-loaded TL reported in Table \ref{tab:CsScintOutput}.  If we instead assume that the fractional energy resolution is inversely proportional to the square root of the energy as given in Reference \cite{Nakamura07, Ejiri2004}, we get:
\begin{equation}
\frac{\sigma(\mbox{Q}_{\beta\beta})}{\mbox{Q}_{\beta\beta}} = 10.8^{+2.0}_{-1.6}\% \times \sqrt{\frac{477\ \mbox{keV}}{3034\ \mbox{keV}}} = 4.3^{+0.8}_{-0.6}\%.
\end{equation}
Furthermore, we observe that the energy resolution of the Mo-loaded scintillator, as compared with the home-made pure scintillator, scales in proportion to the square root of the light output, within uncertainties.  The ratio of the square root of the peak energies and the ratio of the energy resolutions from the \nuc{241}{Am} data (in Table \ref{tab:AmScintOutput}) is:
\begin{equation}
\sqrt{\frac{78.9 \pm 15.8}{42.6 \pm 16.0}} = 1.36 \pm 0.29,\ \ \frac{21.0 \pm 13.4}{16.1 \pm 9.7} = 1.30 \pm 1.14.
\end{equation}
Similarly, the ratio of the square root of the Compton edge energies and the ratio of the energy resolutions from the \nuc{137}{Cs} data (in Table \ref{tab:CsScintOutput}) is:
\begin{equation}
\sqrt{\frac{778^{+6}_{-15}}{498^{+9}_{-10}}} = 1.24^{+0.01}_{-0.02},\ \ \frac{68^{+4}_{-9}}{54^{+10}_{-8}} = 1.26^{+0.24}_{-0.25}.
\end{equation}
This indicates that a single causative factor is responsible for the loss of light, and it augurs well for the energy resolution of scintillator loading recipes that achieve reduced light loss.

\section{Conclusions and Future Work}\label{sec:Concl}
The question of whether the Mo-loaded liquid scintillator we tested could be brought to a level at which it was adequate in terms of light output and energy resolution remains largely an open one.    The Mo-loaded TL had roughly half the light output of commercial scintillators, and an energy resolution  between 4.3 and 4.9\% at \QBB\ (depending on how the extrapolation from lower energy spectral features is done).  This extrapolation is necessary because the highest energy at which we were able to extract the energy resolution of our scintillators was only $\sim 16$\% of \QBB\ (477.3 keV compared to 3034.4 keV) and introduces a large uncertainty.  The light output is adversely affected by Mo loading, which was in saturation for our study.  The resolution, however, just scales as the square root of the light output, and is evidently not degraded by effects other than the light loss.  In future work it would be important to determine the attenuation, an important factor in the design of a large scintillation detector.  Such a detector based on liquid scintillator would probably be designed around a molybdenum concentration of something like one third to one half saturation levels.  Light output and energy resolution would be better at these concentrations than at saturation.  Careful understanding of light transmission and stability at several molybdenum concentrations would be very important in the design of such a detector and could dramatically influence its physics reach.

All in all, our extrapolated energy resolution of 4.3--4.9\% is quite encouraging, given that no effort was made in this study to optimize the fluor concentration or light collection efficiency.  This is of the same order as the NEMO III detector\cite{NEMO3Detector} but larger by a factor 2 than the energy resolutions of ELEGANTS V and the MOON plastic scintillator prototypes \cite{Nakamura07,Ejiri2004,ELEGANTSV,Ejiri2001}.  We note the position dependence of the attenuation results in position dependence of the energy scale. Thus, they must be corrected for, as in case of the MOON plastic scintillator prototypes, to avoid degradation of the resolution in a large volume of liquid scintillator.  A program to optimize the inert chemical environment during scintillator preparation, fluor mixture, and light collection (as well as molybdenum concentration) might yield further improvements on the performance demonstrated here.  The energy resolution will depend not only on the scintillator itself, but also on the shape and size of the scintillator as well as the PMT arrangement.  Additionally, the SNO+ collaboration has made great strides in metal-loading scintillators with neodymium\cite{SNOPlusICHEP2008, SNOPlusMEDEX2007, SNOPlusErice2005}.  SNO+ has focused these efforts on linear alkyl benzene (LAB), which is much easier to handle than TL.  The addition of LAB to future studies of Mo-loaded scintillators would be wise.

There are a number of changes to the program presented in this article that would allow for more definitive understanding of Mo-loaded liquid scintillators.  First, the scintillators would need to be characterized at energies more comparable to \QBB.  This would require a larger volume of scintillator to increase the full-energy $\gamma$-ray collection efficiency.  Volumes approaching one liter would probably be sufficient.  Scintillator characterization using full-energy photo-peaks, rather than Compton edges would further enhance our understanding of scintillator performance.  As stated at the end of Section \ref{sec:ScintLight}, understanding the energy resolution of a Compton edge is rather difficult, and the errors on the resolution presented in Table \ref{tab:CsScintOutput} probably underestimate the actual uncertainty in the parameters extracted from the Compton edges.  Fitting to photo-peaks would allow for both higher energy spectral features and more reliable fits to the scintillation data.  The use of conversion electrons to characterize the energy scale and resolution would also add much to this R\&D program.  The quartz glass in the scintillation test cell was $\sim 1$ mm in thickness.  This is comparable to the range of $\sim 1$-MeV electrons in glass, so it is possible that there are some counts from the \nuc{137}{Cs} conversion electrons in the spectra presented in Figure \ref{fig:CsSpec}.  Placing a windowless conversion electron source in the scintillator would help in their characterization.

Community-wide interest in MOON and SNO+ clearly shows that the capability to load metals into scintillation detectors is an important one to the field of double-beta decay.  The LENS collaboration plans to use In-loaded liquid scintillators to study p-p solar neutrinos.  The Double Chooz experiment\cite{DoubleChooz} will field Gd-doped scintillators in their search for $\theta_{13}$.  R\&D on metal-loading scintillators is focused on both physical processes (such as the interleaving of thin foils between solid plates in the MOON prototype and NEMO detectors), and chemical processes (like the dissolution of Nd in LAB for SNO+).  \nuc{100}{Mo} is an important part of the global \BB\ program, and the chemical dissolution of Mo compounds directly in liquid scintillator would be a tool to further the understanding of \BBz\ decay, astrophysical neutrinos\cite{Ejiri00, Ejiri2002}, and physics beyond the standard Model.

\section{Acknowledgments}
The authors would like to thank Steve Elliott, Andrew Hime, Masaharu Nomachi and Mark Greenfield for their support in the preparation of this article.  This work was performed under Department of Energy contract number DE-FG02-97ER41020 and Los Alamos National Laboratory's Laboratory-Directed Research and Development program.  This article is released under report number LA-UR 09-04230.

\bibliographystyle{unsrt}

\end{document}